\documentclass[runningheads]{llncs}
\usepackage[T1]{fontenc}
\usepackage{graphicx}
\usepackage[colorlinks,citecolor=blue,linkcolor=black]{hyperref}
\usepackage{color}

\usepackage{booktabs,multicol,multirow,colortbl}
\usepackage{amsmath}
\DeclareMathOperator*{\argmax}{argmax}
\usepackage{tikz}
\usepackage{array}
\usepackage{varioref}

\begin{document}
\title{Privacy-Preserving Synthetic Educational Data Generation}
\makeatletter
\newcommand{\printfnsymbol}[1]{%
  \textsuperscript{\@fnsymbol{#1}}%
}
\makeatother
%
% \titlerunning{Abbreviated paper title}
% If the paper title is too long for the running head, you can set
% an abbreviated paper title here
%
% \
%
\author{Jill-Jênn Vie\thanks{Equal contribution.}\inst{1} \and Tomas Rigaux\printfnsymbol{1}\inst{1} \and Sein Minn\inst{2}}
\authorrunning{Vie et al.}
% First names are abbreviated in the running head.
% If there are more than two authors, 'et al.' is used.
%
\institute{SODA, Inria Saclay \and
CEDAR, Inria Saclay \\ 1 rue Honoré d'Estienne d'Orves, 91120 Palaiseau\\
\email{\{jill-jenn.vie,tomas.rigaux,sein.minn\}@inria.fr}}
\maketitle              % typeset the header of the contribution
\begin{abstract}
Institutions collect massive learning traces but they may not disclose it for privacy issues. Synthetic data generation opens new opportunities for research in education. In this paper we present a generative model for educational data that can preserve the privacy of participants, and an evaluation framework for comparing synthetic data generators. We show how naive pseudonymization can lead to re-identification threats and suggest techniques to guarantee privacy. We evaluate our method on existing massive educational open datasets.

\end{abstract}

\section{Introduction}

Educational platforms collect massive amounts of data related to human learning. These can be used to personalize education, train AI-assisted learning systems, but using this data may also harm privacy~\cite{holmes2019ethics,berendt2020ai}. The General Data Protection Regulation (GDPR) protects any information relating to an identified or identifiable natural person. GDPR concerns pseudonymized data, i.e. processing ``so that personal data can no longer be attributed to a specific data subject without the use of additional information'' (Art. 4\footnote{\url{https://gdpr-info.eu/art-4-gdpr/}}) but does not concern anonymized data, i.e. ``personal data rendered anonymous in such a manner that the data subject is not or no longer identifiable'' (Recital 26\footnote{\url{https://gdpr-info.eu/recitals/no-26/}}).

Privacy risk is hard to quantify, as an open dataset can be archived indefinitely, open datasets can be combined, and technology for re-identification is improving over time. There have been a huge number of privacy issues after the re-identification of pseudonymized
data~\cite{de2013unique,narayanan2008robust,rocher2019estimating}.
When a movie-streaming service organized a 1-million-dollar data challenge,
some researchers managed, using solely the movie ratings from the pseudonymized dataset, to match IMDb profiles with the zip code of participants in the pseudonymized dataset~\cite{narayanan2008robust}.

In this paper, we are interested in generating truly anonymized educational records: data that does not belong to anybody, but still shares some interesting properties than real datasets, in order to power technology-enhanced learning.
Our contribution is twofold. We first show how we can generate logs of data using
generative models such as Markov chains or neural networks. We also define a way to practically measure re-identification risk and show how naive pseudonymization techniques, such as dropping a set of rows, or renumbering IDs, are not enough to ensure the privacy of participants. One of our methods is easily scalable as it can generate 1 million rows in 3 seconds while preserving utility and respecting privacy.

This study provides opportunities to open more datasets: instead of just releasing simple statistics, institutions and governments could also provide synthetic datasets so that citizens could provide personalized, innovative solutions for preparing for national examinations. This would benefit research communities such as technology enhanced-learning, educational data mining, and learning analytics, as today it is extremely hard for researchers to have access to student data that is considered too sensitive.

% Actually it is even more important because these days huge deep learning models are trained on large bodies of data which may contain sensitive information, and for the sake of privacy there is no randomness during the training so they may remember information such as passwords or credit card numbers. ref

% Membership inference is about guessing who was in the training set. This emanated from financial transactions; health records. In educational settings it may seem less serious; but actually if you are unique from few records; those can be used to extract other information about you from other sources. \cite{shokri2017membership} The problem is even more serious with user input data, i.e. AOL opening search log queries \cite{korolova2009releasing}.

% We want to train a machine learning model to generate a dataset while minimizing information leakage. 

We first review related work, then introduce the task of privacy-preserving synthetic data generation. We then explain our framework for evaluation, the experiments we made on two real educational datasets, and finally discuss our results.

\section{Related work}

%\subsection{Educational data generation}

In order to protect data, mechanisms such as $k$-anonymity have been considered, i.e. processing the data so that any person is indistinguishable from $k - 1$ other ones in a dataset. However, when we consider high dimensional data, such as mobile geolocation data or educational logging data, then few points are enough to make people unique, therefore $k$-anonymity is no longer feasible: \cite{de2013unique} showed that 4 timestamp-location points are needed to uniquely identify 95\% of individual trajectories in a dataset of 1.5M rows. The uniqueness of a user in a dataset was defined by \cite{rocher2019estimating}, which showed that 15 demographic points are enough to re-identify 99.96\% of Americans. $k$-anonymity also has limitations, as sensitive attributes can be inferred either due to a lack of diversity or using external knowledge~\cite{machanavajjhala2007diversity}.

Some educational research communities attach importance to synthetic or simulated data; while others are mainly interested in real data. For example, in psychometrics, the science of measurement, the validity of a student response model is usually both shown on simulated and real data. ``Pseudo-students'' can also be used to test the quality of an instructional design~\cite{van1990two,vanlehn1994applications}. Generative models, recently famous for deep fakes, are mainly encountered in automatic exercise generation~\cite{cable2013authoring}, simulated response patterns, or student performance prediction, rarely for the generation of a whole dataset. There is a trade-off between generating data that is completely fake, and not very useful; and data that is useful, however easy to re-identify. % Dorodchi et al. use DataSynthesizer to generate student data \cite{dorodchi2019using}. However there is one row per user. In this paper we are dealing with sequences.

% Synthetic data has also been used as a potential solution for investigating human learning at scales. It might be difficult to examine due to a lack of ethically collected data. There is a long history in educational research with the usage of synthetic dataset for various scientific and practical purposes. These goals have been put forward as a possible solution as simulated learners research, such as the use of \Todo{}.

A direct identifier is a specific information that references an individual, such as a name, an e-mail address, or an identification number. A quasi-identifier\footnote{\url{https://edps.europa.eu/system/files/2021-04/21-04-27_aepd-edps_anonymisation_en_5.pdf}} is any piece of information, be it a geographical position at a certain time, or even an opinion on some topic, that could be used, possibly in combination with other quasi-identifiers, with the purpose of re-identifying an individual.
% What if those people are unique? The dataset collects only response data. But this can, in the future, be linked with other data.
In this paper, we are interested to illustrate what can be done using only three simple columns: user ID, item ID and outcome, whether the user got a correct attempt on the item. Our approach can naturally be generalized to several columns, by estimating the conditional probability distributions between variables in order to generate new data that respects those distributions. There are several toolkits to do so, based on Bayesian networks, e.g. sdv.dev~\cite{7796926}; however, in most of them, there is no measurement of re-identification risk. % There is a strong distinction between generation of static data; for which each row is independent (usually one row per user), and dynamic data, for which rows are linked, as we will see now.

\paragraph{Item response theory: estimating outcome given user and item parameters} Response models can be used for estimating both the difficulty of exercises in a questionnaire and the latent abilities of examinees. % Adaptive tests such as GMAT rely on IRT. It can also be used for generating new response data.
% The idea is to measure on a training set the difficulty of parameters which can be used to model the ability parameter of other students.
The Rasch model \cite{rasch1961}, also called 1-parameter logistic, is the most famous and simplest item response theory model (we will denote it by IRT). It is used in real-world adaptive tests such as GMAT, and can also be used to generate synthetic response data.
\[ Pr(R_{ij}=1) = \sigma(\theta_i - d_j) \]
\noindent
where $R_{ij}$ is 1 if user $i$ answers item $j$ correctly, $\sigma : x \mapsto 1 / (1 + \exp(-x))$ is the sigmoid function, $\theta_i$ represents the ability parameter of user $i$ and $d_j$ represents the difficulty of item $j$.

% IRT behaves a bit like logistic regression on sparse features~\cite{vie2019knowledge}. It is usually trained using Newton-Raphson's method.

%\Todo{} ?
%Method is similar from the work of~\cite{niznan2015exploring} to compare the performance of models.

%\subsection{Private synthetic data generation}

% \paragraph{Static data} Most examples we encounter in the literature of privacy consider data with one row per user, usually static, which demographic data such as age or group. The idea is to estimate the conditional probability distributions in order to generate new data that respects those distributions. There are several toolkits based on Bayesian networks such as D

%ith  there are several actions observed for one user, and the process is harder to generate,
%as it should consider a dynamic model.

\paragraph{Privacy-preserving, one row per user} Differential privacy~\cite{dwork2008differential} (DP) is a theoretical framework for proving that the output of a generative model will be indistinguishable by a parameter $\varepsilon > 0$ had a user be present or absent in the training data. It is hard to know which value of epsilon is needed \cite{lee2011much}, but it is related to the budget of queries we can make to the generative model. DP usually relies on adding noise to model weights and is useful for performing queries with privacy guarantees such as histograms~\cite{acs2012differentially}, $n$-grams statistics~\cite{chen2012differentially}. More rarely, DP has been applied to privacy-preserving data generation, usually in settings where there is only one record per user. This is why privacy-preserving Bayesian networks have been proposed such as PrivBayes~\cite{zhang2017privbayes}, implemented in the Python package DataSynthesizer~\cite{ping2017datasynthesizer}. In~\cite{dorodchi2019using}, DataSynthesizer is illustrated on real educational data.

\paragraph{Several rows per user} In our setting, we have several records per user, and we are dealing with the interaction of two entities, users and exercises, that we don't want to protect equally. We want to protect user data, but we want to be able to precisely estimate exercise difficulty. If we were just adding noise to IRT parameters, we would be blurring the utility of our item bank. % Most examples encountered in the privacy literature consider data with one row per user, usually static, with demographic data such as age or group. 
Once we move to high-dimensional scenarios, such as time series, or logging data at irregular time intervals, there are several observations available for each user, and this may arbitrarily increase the risk of re-identification. For example, \cite{leinonen2017preventing} is collecting typing data in order to predict programming experience. They show that delay between keystrokes is enough to re-identify people, but by rounding or bucketing those values, they can still achieve good prediction for the task at hand while reducing re-identification. If the blur is not big enough, people can still be re-identified \cite{de2013unique}.

% With dynamic data, the process is harder to generate, as it should consider a dynamic model.

% Once there are several pieces of information per user, it becomes easier to re-identify, as we will see.

% Most synthetic data generation systems such as  or the sdv.dev toolkit work on static data. We use a graphical network \textcolor{blue}{SM:It is ambiguous: Bayesian network is also a graphical network} to decide in which order to generate which cells.

% From which we get inspired. Some people have to generate fake data. Others have to reidentify. It is similar to how generative adversarial networks work. But with different teams.

% We can use blur, rounding, or bucketing.

\begin{table}[ht]
\caption{Example of minimal tabular dataset.}
\label{example-dataset}
\centering
\begin{tabular}{ccc} \toprule
user ID & action ID & outcome \\ \midrule
2487 & 384 & 1 \\
2487 & 242 & 0 \\
2487 & 39 & 1 \\
2487 & 65 & 1 \\ \bottomrule
\end{tabular}
\arrayrulecolor{white}
\begin{tabular}{l} \toprule
description \\ \midrule
user 2487 got token ``I'' correct \\
user 2487 got token ``ate'' incorrect \\
user 2487 got token ``an'' correct \\
user 2487 got token ``apple'' correct \\ \bottomrule
\end{tabular}
\arrayrulecolor{black}
\end{table}

\section{Privacy-Preserving Synthetic Data Generation}

\subsection{Goal}

A synthetic dataset should have several properties:

\paragraph{Utility} The fake dataset should bear a strong similarity to the real dataset (histograms, similar results to queries). Also if we conduct a study, e.g. estimating item difficulties using an IRT model, the learned parameters should be similar for both the real and the generated dataset.
\paragraph{Privacy} It should not be easy to re-identify participants in the real dataset from the synthetic dataset.\bigskip

For example, it is easy to generate random noise, and complete dummy datasets with guaranteed privacy, but it won't be useful if we do not preserve correlation between columns.

For the sake of simplicity, we assume that the data is provided as triplets $(i, j_t, r_t) = (\textsf{userID}, \textsf{actionID},
\textsf{outcome})$ where the \textsf{outcome} $r_t$ is 1 if user $i$ makes a successful action $j_t$ and 0 otherwise. See Table~\ref{example-dataset} for an example of such dataset.

We first need a model of sequence prediction, to identify which action comes next. Formally, we need a model of $p(j_{t + 1}|j_t,\ldots,j_1)$. Then, we need a response model $p(r_t|i,j_t)$.

\subsection{Sequence generation}
\paragraph{Markov chains} This simple probabilistic graphical model has been used for generating text, music, etc.
It relies on a probability transition for jumping from one action to another: $P_{su} = \Pr(j_{t + 1} = u|j_t = s)$ is the probability to jump from action $s$ to action $u$. The Markov chain is trained on existing corpus of actions. Once the $P$ matrix has been estimated, it can be used to sample a random walk from action to action. A Markov chain is said memoryless because the next action only depends on the current action: $p(j_{t + 1}|j_t,\ldots,j_1) = p(j_{t + 1}|j_t)$.

% Markov chains or their variants are also used for sequence prediction: bayesian knowledge tracing \cite{corbett1994knowledge} is a hidden Markov model, i.e. the latent state is jumping from no knowledge to knowledge \textcolor{blue}{SM:it is also ambiguous, (the knowledge mastery) latent state is increased or decreased gradually}; and the observed variable is the outcome. In knowledge tracing, we want to know, given the successful or unsuccessful outcomes of users over exercises, whether they will get correct the next exercise.

\paragraph{Recurrent neural networks} Neural networks are famous for natural language processing, and generation. RNNs have been used in knowledge tracing for predicting student performance \cite{piech2015deep}.
They have many more parameters, so they can remember more than simple Markov chains, but they are way slower to train. Some works have shown that
a simple updated IRT model could match the performance of RNN~\cite{wilson2016back} for knowledge tracing. \cite{gervet2020deep} has shown that it depends on how much the dataset contains long sequences and if the sequential aspect of the dataset is prominent. A Gated recurrent unit (GRU) is an example of RNN. In our case, the input is sequence $(j_1, \ldots, j_t)$ and the output is sequence $(\widehat{j}_2, \ldots, \widehat{j}_{t + 1})$ and GRU computes:%
\begin{align*}
\begin{aligned}[c]
    s_t &= \sigma\left(W \textcolor{red}{j_t} + U h_{t-1} + b\right) \\
    z_t &= \sigma\left(W' \textcolor{red}{j_t} + U' h_{t-1} + b'\right)
\end{aligned} \qquad
\begin{aligned}[c]
    \hat{h}_t &= \tanh\left(W'' + U'' (s_t * h_{t-1}) + b''\right) \\
    h_t &= (1 - z_t) * h_{t - 1} + z_t * \hat{h}_t \qquad h_0 = 0\\
    \textcolor{red}{\widehat{j}_{t + 1}} &= \argmax(W''' h_t + b''') 
\end{aligned}
\end{align*}

\noindent
where $\sigma$ is the same sigmoid function as in IRT, $*$ denotes element-wise product, and parts of input and output where shown in red for clarity.

\subsection{Response pattern generation}

Once the sequence of skills has been generated, what is left is to generate outcomes. For this we use the Rasch model:
\[ p(r_t = 1|i,j_t) = \sigma(\theta_i - d_{j_t}). \]

\begin{figure}[ht]
    \centering
    \includegraphics[width=5cm]{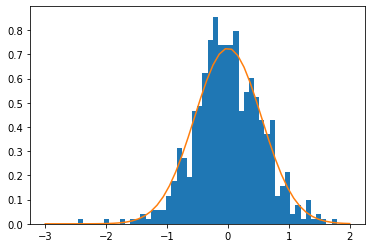}
    \caption{From the estimated $\theta$ parameters from the training set, it is easy to fit a Gaussian and sample new users from it.}
    \label{fit-gaussian}
\end{figure}

We fit an IRT model on the training dataset to learn the $\theta_i$ ability of each user $i$ and the difficulty $d_j$ of each action $j$. Then, to generate new users, we just need to fit a normal distribution on the histogram of existing $\theta$ values and sample from it to generate responses using the IRT model and the estimated action difficulties $d_j$, see Figure~\ref{fit-gaussian}. This is the core of our strategy: as the generated $j_t$ and the sampled $\theta$ do not correspond to any particular user anymore, the generated dataset should be anonymous.

\section{Evaluation Framework}

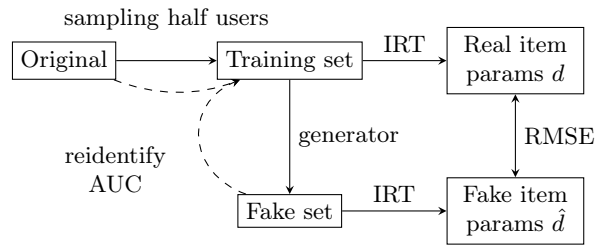
\begin{figure}[hb]
    \centering
    \begin{tikzpicture}[
    xscale=3,
    yscale=2,
    data/.style={draw},
    >=stealth
]
\node[data] (original) at (0,0) {Original};
\node[data] (training) at (1,0) {Training set};
\node[data] (fake) at (1,-1) {Fake set};
\node[data,text width=1.6cm,text centered] (real-irt) at (2,0) {Real item params $d$};
\node[data,text width=1.6cm,text centered] (fake-irt) at (2,-1) {Fake item params $\hat{d}$};
\draw[->] (original) edge node[above=3mm] {sampling half users} (training);
\draw[->] (training) edge node[right] {generator} (fake);
\draw[<->] (real-irt) edge node[right] {RMSE} (fake-irt);
\draw[->,dashed,bend right] (original) edge (training);
\draw[->,dashed,bend left=60,text width=2cm,text centered] (fake) edge node[below left] {reidentify\\AUC} (training);
\draw[->] (training) edge node[above] {IRT} (real-irt);
\draw[->] (fake) edge node[above] {IRT} (fake-irt);
\end{tikzpicture}
    \caption{The architecture of our study}
    \label{archi}
\end{figure}

To compare strategies for educational data generation, our architecture is described in Figure~\ref{archi} and explained in this Section.

\subsection{Training set sampling and generation}
\label{training-set-sampling}

For each original dataset, we first sample a training set that will be used to train the generators. This training set contains the rows that belong to half of all users. Then our models will generate new, synthetic (or fake) tabular datasets.

\subsection{Utility}

To compare the real and fake sets, we first compute some histograms for the real and generated sequences:
a number of occurrences of each action, sequence lengths, and distribution of repeated skills.

Once the fake dataset has been generated, we want to know whether training an IRT model to estimate the difficulty of actions has similar findings on the real dataset and on the fake dataset. We compute the root mean squared error of action difficulty parameters learned by IRT between the training set and the fake set. The weighted RMSE (denoted $wRMSE$) is given by the following formula:

\[ wRMSE = \sqrt{\sum_{i = 1}^N w_i (d_i - \hat{d}_i)^2} \]

The usual RMSE is when all actions are equally weighted, i.e. $w_i = 1/N$ for all $i$. However some actions are less frequent than others, so it is normal that their parameter is not well estimated. Therefore we also introduce a weighted RMSE where $w_i$ corresponds to the frequency of action $i$ in the training set, i.e. its number of occurrences divided by the size of the training set.

% For the Assistments dataset we also compare the histogram of repeated skills. For the Duolingo we don't do it because it is mainly 1: there are few consecutive repeats.

\subsection{Reidentification score}

As a measure of how easy it is to re-identify people, we borrow the practical task of membership inference encountered in \cite{jordon2020hide,shokri2017membership}. We assume that an adversary has access to the original dataset (e.g. some auxiliary information about the population from the outside world) and the fake generated dataset, and wants to guess which users were in the training set. This is a classification problem where for each user in the original dataset, we want to guess 1 if it was present in the training set used to generate the fake set, and 0 otherwise. We will now give examples of why membership inference is already an issue: if the dataset used for training the fake set corresponds to some query, i.e. ``students with special needs'' or ``students having a certain socioeconomic status'', then membership inference is already something that may harm privacy. More generally, if one person can be re-identified just by a few actions, then using other sources of information (e.g. cookies, other databases), these actions can be used to uniquely describe this user, and re-identify them in other databases. This is exactly the example of the Netflix Prize~\cite{narayanan2008robust}.

Once a classifier is performing membership inference, its performance can be evaluated using the area under the ROC curve (AUC), a number between 0 and 1. Any random guess should have an AUC of 0.5, as half of the original people belong to the training set, as stated in Section~\ref{training-set-sampling}.

% It would help identifying which university did these students attend to, based only on their actions.

% From how many data points is a person unique?

% To check that, we select randomly \(n\) points from a sequence. We check

\section{Experiments}

\subsection{Datasets}

Datasets are described below and their statistics are reported in Table~\ref{data-stats}. Median refers to the median across all users of the median number of repeats for each skill. Max refers to the maximum across all users of the maximum number of repeats for a skill.

\paragraph{Assistments 2009} This dataset contains 279,000 outcomes of 4,163 students attempting math exercises. Each exercise is mapped to one among 112 knowledge components in Mathematics \cite{heffernan2014assistments}. This dataset is popular in the Educational Data Mining community, notably for knowledge tracing. Here, we consider that a skill corresponds to an action.

\paragraph{Duolingo 2018} This massive dataset contains the outcomes of 1,213 English-speaking people learning French. It contains 1.2M logs of users attempting to type words in the Duolingo app. The actions are the words expected in the correct answer, and the outcomes are at the word level: 1 for a correctly typed word, 0 for a spelling mistake, see again Table~\vref{example-dataset} for an example. This dataset was part of the Duolingo competition at NAACL-HLT 2018 for a knowledge tracing task \cite{settles2018second}.\bigskip

We remove actions where the success rate is either 0\% or 100\%, as those are either too easy or impossible to get right, and their corresponding IRT parameters are $\pm \infty$. For example, in the Duolingo dataset, the French word « train » had 0\% success rate. We were surprised so we looked at the expected sentences and discovered that it was in fact the « en train de » locution, which is the translation of the -ing form in English, which is hard to get right for English people learning French (``She is eating'' $\leftrightarrow$ « Elle est en train de manger »).

\begin{table}
    \caption{Statistics for the datasets considered in the study.}
    \centering
    \begin{tabular}{|c|ccc|>{\centering}p{1.3cm}>{\centering}p{1.cm}|>{\centering}p{1.3cm}>{\centering}p{1.cm}>{\centering}p{1.cm}|} %\toprule
    \hline
\multirow{2}{*}{Dataset} & \multirow{2}{*}{Size} & \multirow{2}{*}{Users} & \multirow{2}{*}{Actions} & \multicolumn{2}{c|}{Repeated actions} & \multicolumn{3}{c|}{Sequence length}\tabularnewline
& & & & Med & Max & Min & Med & Max\tabularnewline  \hline %\midrule
        Assistments 2009 & 279k & 4163 & 112 & 3 & 144 & 1 & 20 & 1021 \tabularnewline
        Duolingo SLAM 2018 & 1.2M & 1213 & 2416 & 1 & 4 & 90 & 742 & 10008 \tabularnewline %\bottomrule
        \hline
    \end{tabular}
    \label{data-stats}
\end{table}

\subsection{Generative models}

\paragraph{Baseline Drop} As a baseline, we drop a certain amount of rows from the training set (a ratio $r \in \{0., 0.25, 0.5, 0.75, 0.99, 0.999\}$), then randomize user IDs.

% Or we combine some small sequences to make bigger sequences. \Todo for another time

\paragraph{Markov chain} The Markov chain for generating actions was implemented using the \texttt{lea} Python package for discrete probability distributions~\cite{denis2020probabilistic}. As parameters, we define a length limit of 1000 for Assistments and 10000 for Duolingo. Our Markov chain takes 3 seconds to train and generate the Duolingo dataset.

\paragraph{RNN} Our recurrent neural network is a Gated Recurrent Unit (GRU) implemented in PyTorch. The batch size was 64 for Assistments and 16 for Duolingo. We minimize the cross entropy loss of observed actions using the Adam optimizer~\cite{kingma2014adam}. Training takes approximately two hours for the Duolingo dataset. It is trained on smaller sequences first then longer sequences.

\paragraph{IRT} For generating the outcomes from user parameters and actions, we use a Rasch model denoted by IRT, implemented as \texttt{LogisticRegression} in the scikit-learn package \cite{pedregosa2011scikit}. We use the default regularization parameter $C=1$.

\subsection{Re-identification model}

We compute, for each original sequence, the longest common subsequence with each fake sequence in the fake dataset. This is performed in $O(\ell \ell')$ time for a pair of sequences of lengths $\ell$ and $\ell'$, so $O(MN)$ in total where $M$ is the size of the original dataset and $N$ the size of the fake dataset. Our implementation is written in C++.

Then we take the maximum of those matching scores divided by the length of the fake sequence, i.e. best-normalized percentage of matching. It gives a matching score for each original user, used for the classification task of membership inference. The quality of re-identification is estimated using AUC.

We limit the re-identification to users with enough information. More precisely, we define the cumulative entropy of a user as $\sum_t-p(j_t) \log p(j_t)$, where ${\left(j_t\right)}_t$ is its sequence of actions and $p(j)$ the frequency of action $j$ in the original dataset.
We then only try to re-identify users with an entropy larger than $-p \log p$ for $p$ the proportion of users in the training dataset ($p=0.5$ in our experiments), i.e. the entropy of the information ``user as part of the training dataset''. In Assistments this induces filtering of 15\% of users while in Duolingo it does not change anything, as sequences are already pretty long and diverse, so they contain a lot of information already.

Our experiments can be reproduced using our code which is free and open source software\footnote{\url{https://github.com/Akulen/PrivGen}}.

\section{Results and Discussion}

\subsection{First look at the synthetic datasets}

\begin{figure}
    \centering
    \includegraphics[width=0.4\linewidth]{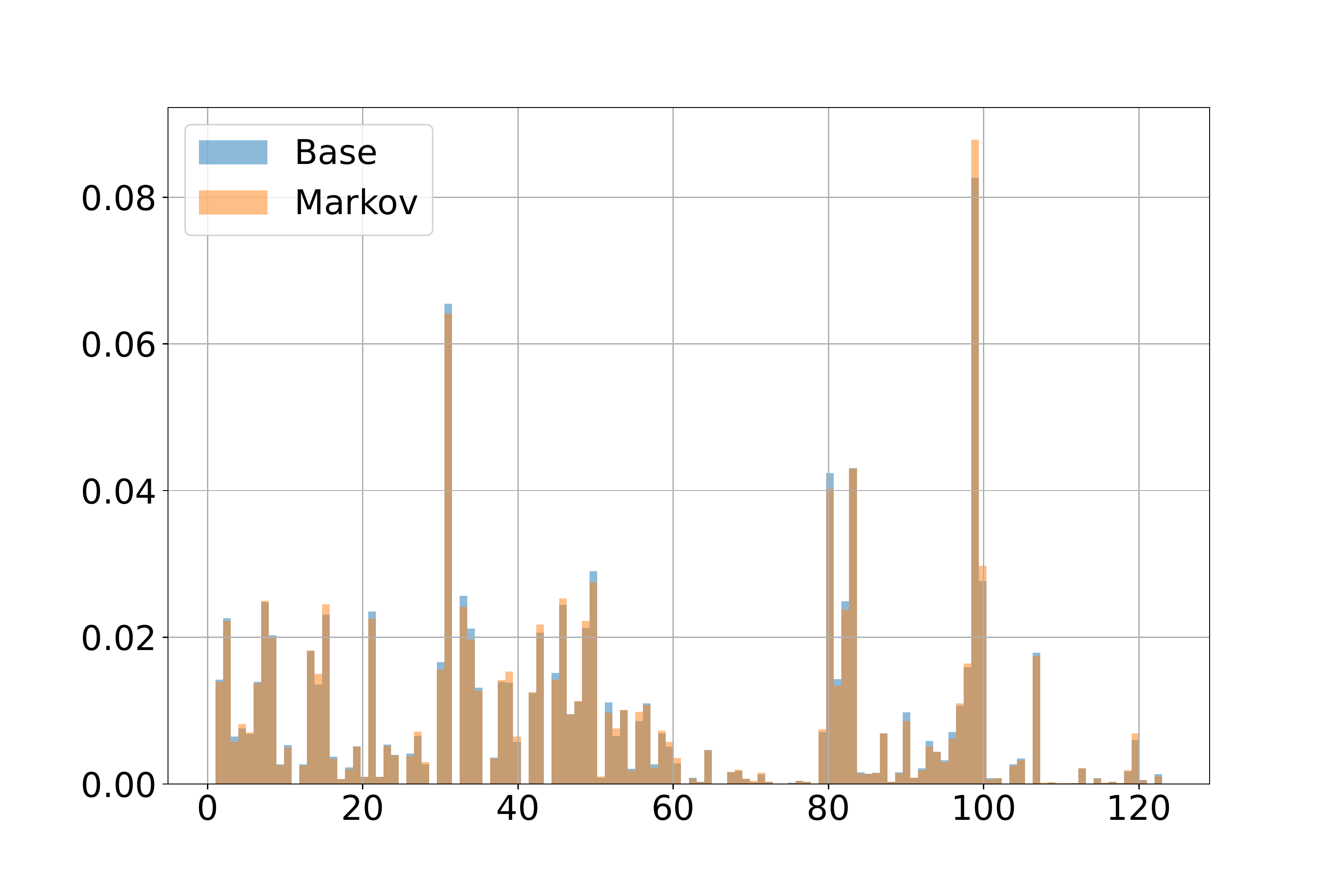}
    \includegraphics[width=0.4\linewidth]{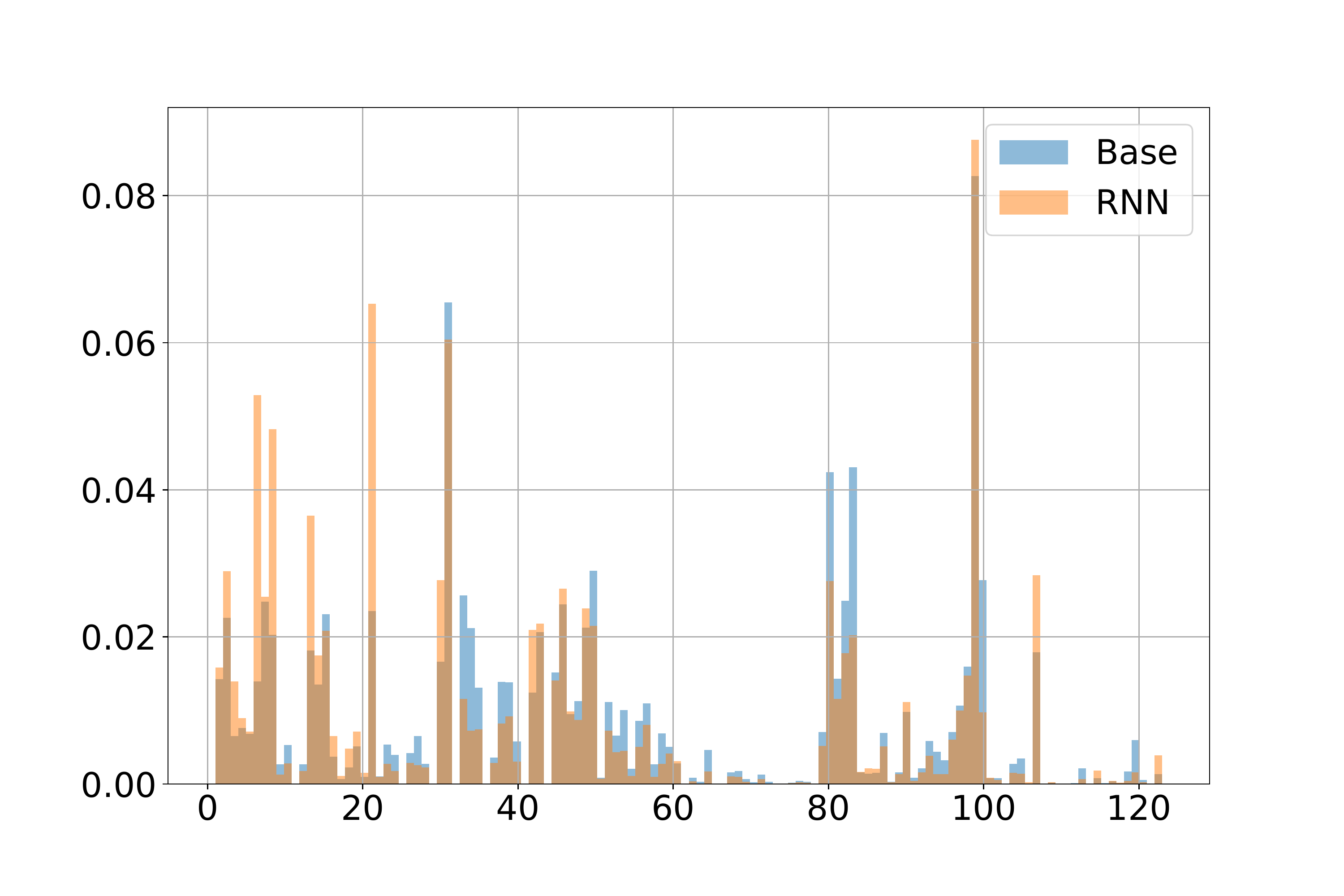}
    \caption{Histogram of actions for the original and generated datasets by Markov chain and RNN.}
    \label{action-hist}
\end{figure}

We first compare the histogram of actions in Figure~\ref{action-hist}, where Base represents the training set. We see that the Markov chain, a very simple model, approximates the skill histograms better than RNN.

% Figure~\ref{ass09conc} represents the log counts of occurrences of repeated actions, for the training and generated sets. RNN generates 7 sequences with too many repeated actions, once those are removed, it is closer to the original distribution than Markov. Probably this is due to the memoryless property of the Markov chain.

% \begin{figure}
%     \centering
%     \includegraphics[width=5cm]{figures/assistments2009_concentration.png}
%     \caption{Log counts of occurrences of repeated actions in the Assistments 2009 dataset.}
%     \label{ass09conc}
% \end{figure}

We give examples of generated sequences from our approach on the Duolingo dataset in Table~\ref{example-gen-sentences}. For clarity, we do not display the outcomes, only actions which are French words. Markov chains are simple, but it is a memoryless process that explains why bigrams (consecutive words) are preserved but not whole sentences (e.g. « Il faut du fromage et juin à midi »). RNNs are longer to train but they can preserve longer contexts, such as generating several sentences in the same theme (food for the second sentence, or animals in the third sentence), like in the original dataset. However, it may not preserve bigrams (e.g. « des robe »).

\begin{table}[ht]
\caption{Original vs. generated Duolingo sequences.}
\resizebox{\textwidth}{!}{%
\begin{tabular}{ll} \toprule
\multirow{4}{*}{Original} & 1. La femme Je suis rouge L' homme Je suis riche Je mange Il est riche Je suis calme\\
    & 2. Je suis riche Je suis rouge L' homme Je mange Il est riche Je suis calme ...\\
    & 3. Je suis rouge Je suis riche L' homme Il est riche Je mange Je suis calme ...\\
    & 4. ... Les chiens Les chiens Vous êtes grand Je mange des baguettes\\ \midrule
\multirow{5}{*}{\parbox{1.5cm}{Markov\\ generated}} & 1. Le costume La bière est rouge Les filles mangent Cet homme est riche\\
 & 2. Aux mois d' accord Tu es grande Je parle Qui suis riche L' éléphant Ma femme\\
 & 3. Le tigre Le menu Le sac est un costume Quoi Combien Oui je vais bien\\
 & 4. Quatorze enfants C' est violet Ma robe Il faut du fromage et juin à midi\\
 & Vous avez un animal Vous mangez une secrétaire Sinon je sais\\
 & Le cheval gagne Ça va Oui je sais Je motive mon chien\\ \midrule
 \multirow{6}{*}{\parbox{1.5cm}{RNN\\ generated}} & 1. Nous mangeons Nous apprenons Je parle Il parle Je parle Je sais Il faut\\
 & J' aime le fromage Je veux veux un poisson\\
 & 2. Le bonbon est rouge J' aime boire La carotte J' aime manger Un œuf\\
 & La confiture Je bois une boisson rouge\\
 & 3. Tu es en train de manger Un dauphin Le chat est noir Le éléphant est vert\\
 & 4. Il faut du pain Elle pose des chats Les chiennes Il pleut des frites\\
 & Ces enfants mangent des robe
 \\ \bottomrule
\end{tabular}}
\label{example-gen-sentences}
\end{table}

\subsection{Quality and re-identification trade-off}

\begin{table}[ht]
\caption{Results. RMSE should be low for good utility and Re-identification AUC should be low for good privacy. The best results are shown in bold.}
\centering
\begin{tabular}{|c|c|cccccc|c|c|} \hline %\toprule
\multirow{2}{*}{Dataset} & \multirow{2}{*}{Metric} & \multicolumn{6}{c|}{Drop} & \multirow{2}{*}{MC} & \multirow{2}{*}{RNN} \\
& & 0. & 0.25 & 0.5 & 0.75 & 0.99 & 0.999 & &\\ \hline %\midrule
\multirow{3}{*}{\parbox{2.5cm}{\centering ASSISTments\\ 2009}} & RMSE & \textbf{0.000} & \textbf{0.093} & \textbf{0.147} & 0.283 & 0.719 & 0.833 & \textbf{0.245} & \textbf{0.213} \\
& wRMSE & \textbf{0.000} & \textbf{0.035} & \textbf{0.064} & 0.105 & 0.481 & 0.692 & \textbf{0.065} & \textbf{0.061} \\
& Re-ID AUC & 0.913 & 0.776 & 0.680 & 0.588 & \textbf{0.497} & \textbf{0.497} & \textbf{0.495} & \textbf{0.508} \\ \hline
\multirow{3}{*}{\parbox{2.5cm}{\centering Duolingo SLAM\\ 2018}} & RMSE & \textbf{0.000} & \textbf{0.208} & \textbf{0.308} & 0.450 & 0.730 & 0.793 & \textbf{0.369} & 0.431 \\
& wRMSE & \textbf{0.000} & \textbf{0.067} & \textbf{0.113} & 0.195 & 0.624 & 0.985 & \textbf{0.114} & \textbf{0.143} \\
& Re-ID AUC & 1.000 & 1.000 & 1.000 & 1.000 & 0.554 & \textbf{0.506} & \textbf{0.511} & \textbf{0.516} \\ \hline %\bottomrule
\end{tabular}
\label{auc-rmse}
\end{table}

\begin{figure}[ht]
    \centering
    \includegraphics[width=0.49\linewidth]{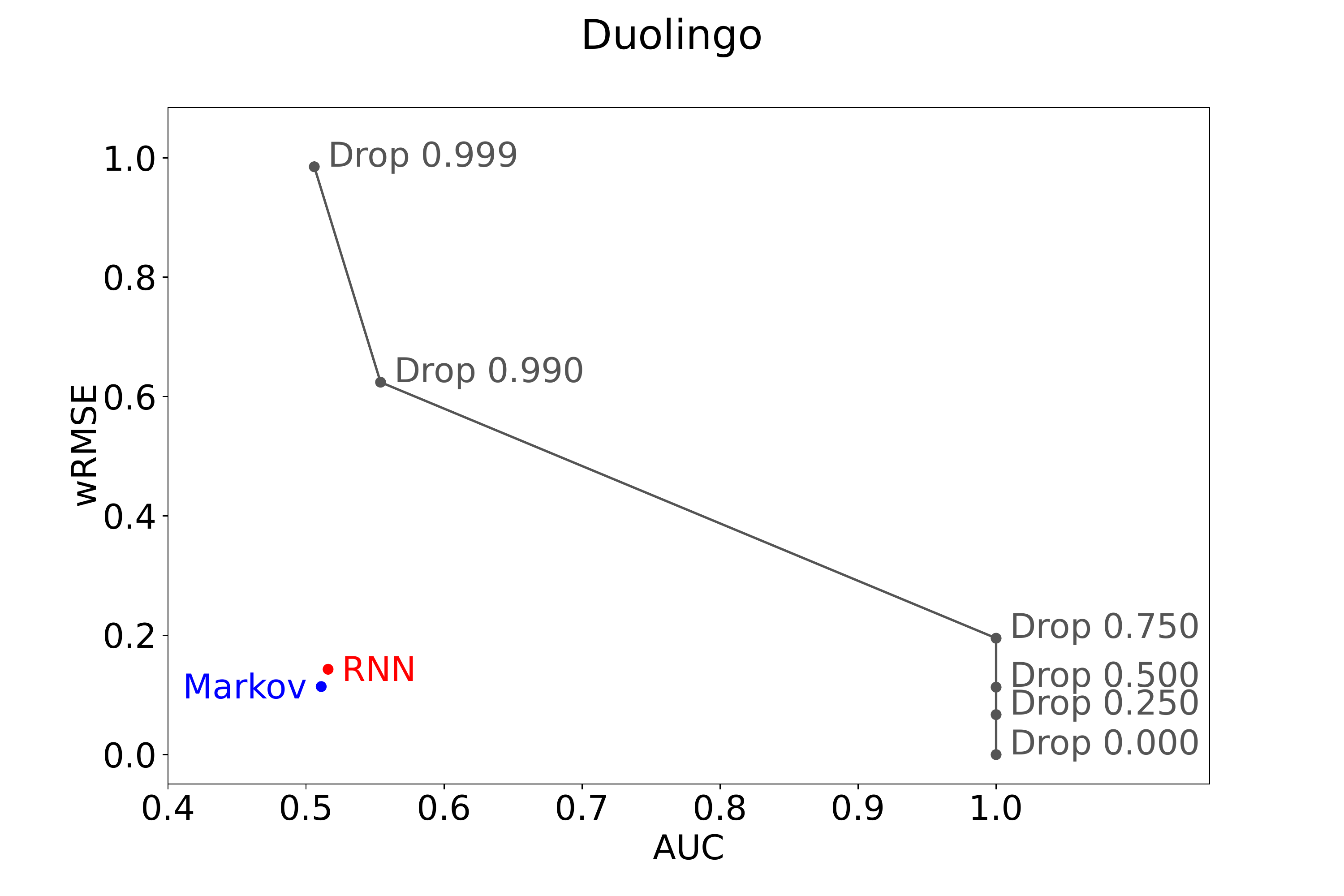}
    \includegraphics[width=0.49\linewidth]{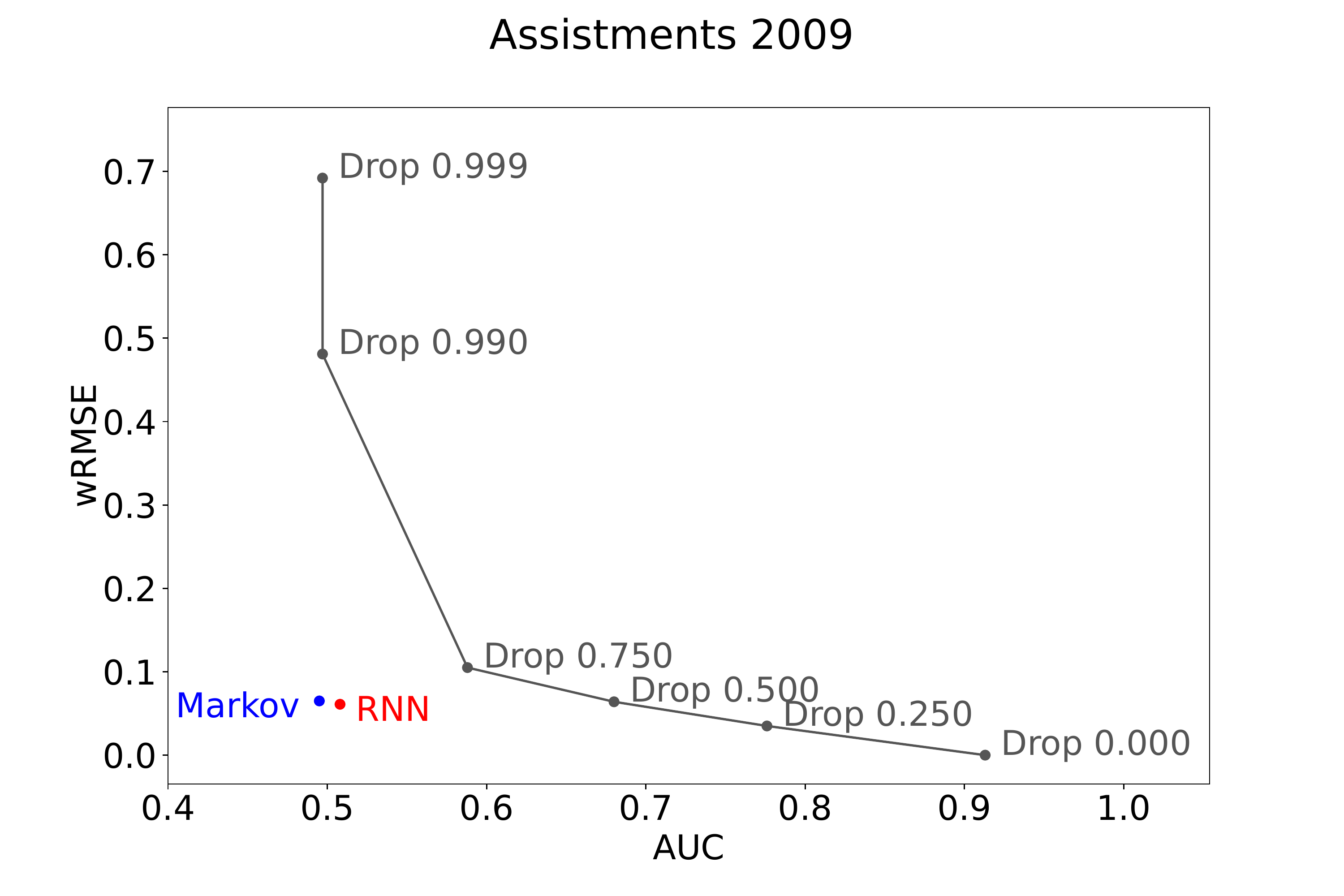}
    \caption{The trade-off between quality (low weighted RMSE) and privacy (low re-identification AUC) for all models considered. The bottom left is better.}
    \label{auc-rmse-duolingo}
\end{figure}

Quantitative results are provided in Table~\ref{auc-rmse} and Figure~\ref{auc-rmse-duolingo}. Drop baselines with ratios of 0.5 and 0.75 are the worst of possible worlds: loss in estimation quality of the action difficulty parameters, and easy membership inference. What is quite remarkable is that on the Duolingo dataset, even if we drop 75\% of rows, it is still possible to exactly recover 100\% of the training set. This is probably because there are more tokens and sequences are longer (the minimal length is 90 and the median length is 742), so people are more easily unique.

Markov chain and RNN have comparable quality RMSE scores to the Drop baseline for low ratio. But even Drop 0, which corresponds to keeping all lines and rewriting the user IDs, is very easy to re-identify (AUC 0.913), which shows that simple pseudonymization is not enough. Therefore, the best models are Markov chain and RNN, which is particularly visible in Figure~\ref{auc-rmse-duolingo}. This means we can freely share the fake dataset: it will follow a similar distribution to the real one, but the underlying ``users'' do not exist; they cannot be re-identified.

\section{Limitations, impact and future work}

A limitation is that so far we consider a model of evolution for the skill $j$, i.e. the question that is assessed at each time, but not for the user ability $\theta$, i.e. a learning model. Natural extensions would be to consider knowledge tracing models such as PFA~\cite{pavlik2009performance} or more sophisticated ones such as DAS3H~\cite{choffin2019das3h}, to get dynamic models of learning.

In future work, we'd like to test our setting on more sophisticated tabular datasets: users would be even more unique. We notably want to work on timestamps, as the delays between attempts may be unique between participants, therefore may harm privacy. In this paper, we were interested in learning item difficulties, but other applications may have a different objective to optimize. We want to highlight the fact that for the sake of researchers in technology-enhanced learning, item parameters should be as open as possible; while for the sake of students, user parameters should be kept as private as possible.

The example shown in this paper helps raise awareness in what can be done with student data. Our re-identification task of membership inference may seem a bit weak, so here is a more precise example. Let us now assume that for the sake of providing accurate recommendations, a dataset of student logs with a particular condition, say ADHD, is shared. We show that it could be possible, having access to a bigger dataset of students logs, to identify which students have ADHD. Personalized education should be able to provide further help to students with special needs, without letting anyone know which student has what condition.

\section{Conclusion}

In this paper, we show how we can generate educational data records for research while preserving
the privacy of real users. We illustrated that naive pseudonymization or dropping rows from a dataset is not enough, as techniques based on text mining can re-identify who was in the training set. Our approach generates fake users, thus anonymized data that can be freely shared. We advocate for more open datasets to nurture educational research and foster technology-enhanced learning; but privacy-preserving, synthetically generated ones.

% Links can be made between algorithms to reidentify music~\cite{wang2003industrial} or bird songs using bit vectors; and algorithms to re-identify humans in tabular datasets.

%Dropping a set of rows still reveals entries that correspond to actual people. We show how longest common subsequence and longest common substring methods can re-identify who is the most plausible record our subset may correspond to.

\bibliographystyle{splncs04}
\bibliography{biblio}

\begin{thebibliography}{10}
\providecommand{\url}[1]{\texttt{#1}}
\providecommand{\urlprefix}{URL }
\providecommand{\doi}[1]{https://doi.org/#1}

\bibitem{acs2012differentially}
Acs, G., Castelluccia, C., Chen, R.: Differentially private histogram
  publishing through lossy compression. In: 2012 IEEE 12th International
  Conference on Data Mining. pp. 1--10. IEEE (2012)

\bibitem{berendt2020ai}
Berendt, B., Littlejohn, A., Blakemore, M.: Ai in education: learner choice and
  fundamental rights. Learning, Media and Technology  \textbf{45}(3),  312--324
  (2020)

\bibitem{cable2013authoring}
Cabl{\'e}, B., Guin, N., Lefevre, M.: An authoring tool for semi-automatic
  generation of self-assessment exercises. In: International Conference on
  Artificial Intelligence in Education. pp. 679--682. Springer (2013)

\bibitem{chen2012differentially}
Chen, R., Acs, G., Castelluccia, C.: Differentially private sequential data
  publication via variable-length n-grams. In: Proceedings of the 2012 ACM
  conference on Computer and communications security. pp. 638--649 (2012)

\bibitem{choffin2019das3h}
Choffin, B., Popineau, F., Bourda, Y., Vie, J.J.: {DAS3H}: modeling student
  learning and forgetting for optimally scheduling distributed practice of
  skills. arXiv preprint arXiv:1905.06873  (2019)

\bibitem{de2013unique}
De~Montjoye, Y.A., Hidalgo, C.A., Verleysen, M., Blondel, V.D.: Unique in the
  crowd: The privacy bounds of human mobility. Scientific reports
  \textbf{3}(1), ~1--5 (2013)

\bibitem{denis2020probabilistic}
Denis, P.: Probabilistic inference using generators: The statues algorithm. In:
  Science and Information Conference. pp. 133--154. Springer (2020)

\bibitem{dorodchi2019using}
Dorodchi, M., Al-Hossami, E., Benedict, A., Demeter, E.: Using synthetic data
  generators to promote open science in higher education learning analytics.
  In: 2019 IEEE International Conference on Big Data (Big Data). pp.
  4672--4675. IEEE (2019)

\bibitem{dwork2008differential}
Dwork, C.: Differential privacy: A survey of results. In: International
  conference on theory and applications of models of computation. pp. 1--19.
  Springer (2008)

\bibitem{gervet2020deep}
Gervet, T., Koedinger, K., Schneider, J., Mitchell, T., et~al.: When is deep
  learning the best approach to knowledge tracing? Journal of Educational Data
  Mining  \textbf{12}(3),  31--54 (2020)

\bibitem{heffernan2014assistments}
Heffernan, N.T., Heffernan, C.L.: The assistments ecosystem: Building a
  platform that brings scientists and teachers together for minimally invasive
  research on human learning and teaching. International Journal of Artificial
  Intelligence in Education  \textbf{24}(4),  470--497 (2014)

\bibitem{holmes2019ethics}
Holmes, W., Iniesto, F., Sharples, M., Scanlon, E.: {ETHICS in AIED: Who Cares?
  An EC-TEL workshop}. In: EC-TEL 2019 Fourteenth European Conference on
  Technology Enhanced Learning (2019), \url{http://oro.open.ac.uk/67263/}

\bibitem{jordon2020hide}
Jordon, J., Jarrett, D., Yoon, J., Barnes, T., Elbers, P., Thoral, P., Ercole,
  A., Zhang, C., Belgrave, D., van~der Schaar, M.: Hide-and-seek privacy
  challenge. arXiv preprint arXiv:2007.12087  (2020)

\bibitem{kingma2014adam}
Kingma, D.P., Ba, J.: Adam: A method for stochastic optimization. arXiv
  preprint arXiv:1412.6980  (2014)

\bibitem{lee2011much}
Lee, J., Clifton, C.: How much is enough? choosing $\varepsilon$ for
  differential privacy. In: International Conference on Information Security.
  pp. 325--340. Springer (2011)

\bibitem{leinonen2017preventing}
Leinonen, J., Ihantola, P., Hellas, A.: Preventing keystroke based
  identification in open data sets. In: Proceedings of the Fourth (2017) ACM
  Conference on Learning@Scale. pp. 101--109 (2017)

\bibitem{machanavajjhala2007diversity}
Machanavajjhala, A., Kifer, D., Gehrke, J., Venkitasubramaniam, M.:
  l-diversity: Privacy beyond k-anonymity. ACM Transactions on Knowledge
  Discovery from Data (TKDD)  \textbf{1}(1),  3--es (2007)

\bibitem{narayanan2008robust}
Narayanan, A., Shmatikov, V.: Robust de-anonymization of large sparse datasets.
  In: 2008 IEEE Symposium on Security and Privacy (sp 2008). pp. 111--125. IEEE
  (2008)

\bibitem{7796926}
{Patki}, N., {Wedge}, R., {Veeramachaneni}, K.: The synthetic data vault. In:
  2016 IEEE International Conference on Data Science and Advanced Analytics
  (DSAA). pp. 399--410 (Oct 2016). \doi{10.1109/DSAA.2016.49}

\bibitem{pavlik2009performance}
Pavlik~Jr, P.I., Cen, H., Koedinger, K.R.: Performance factors analysis--a new
  alternative to knowledge tracing. Online Submission  (2009)

\bibitem{pedregosa2011scikit}
Pedregosa, F., Varoquaux, G., Gramfort, A., Michel, V., Thirion, B., Grisel,
  O., Blondel, M., Prettenhofer, P., Weiss, R., Dubourg, V., et~al.:
  Scikit-learn: Machine learning in python. the Journal of machine Learning
  research  \textbf{12},  2825--2830 (2011)

\bibitem{piech2015deep}
Piech, C., Bassen, J., Huang, J., Ganguli, S., Sahami, M., Guibas, L.J.,
  Sohl-Dickstein, J.: Deep knowledge tracing. In: Advances in neural
  information processing systems. vol.~28 (2015)

\bibitem{ping2017datasynthesizer}
Ping, H., Stoyanovich, J., Howe, B.: Datasynthesizer: Privacy-preserving
  synthetic datasets. In: Proceedings of the 29th International Conference on
  Scientific and Statistical Database Management. pp.~1--5 (2017)

\bibitem{rasch1961}
Rasch, G.: {On General Laws and the Meaning of Measurement in Psychology}. In:
  Proceedings of the Fourth Berkeley Symposium on Mathematical Statistics and
  Probability, Volume 4: Contributions to Biology and Problems of Medicine. pp.
  321--333. University of California Press, Berkeley, Calif. (1961),
  \url{https://projecteuclid.org/euclid.bsmsp/1200512895}

\bibitem{rocher2019estimating}
Rocher, L., Hendrickx, J.M., De~Montjoye, Y.A.: Estimating the success of
  re-identifications in incomplete datasets using generative models. Nature
  communications  \textbf{10}(1), ~1--9 (2019)

\bibitem{settles2018second}
Settles, B., Brust, C., Gustafson, E., Hagiwara, M., Madnani, N.: Second
  language acquisition modeling. In: Proceedings of the thirteenth workshop on
  innovative use of NLP for building educational applications. pp. 56--65
  (2018)

\bibitem{shokri2017membership}
Shokri, R., Stronati, M., Song, C., Shmatikov, V.: Membership inference attacks
  against machine learning models. In: 2017 IEEE symposium on security and
  privacy (SP). pp. 3--18. IEEE (2017)

\bibitem{van1990two}
Van~Lehn, K.: Two pseudo-students: Applications of machine learning to
  formative evaluation. Tech. rep., Carnegie-Mellon Univ, Pittburgh, PA, Dept
  of Psychology (1990)

\bibitem{vanlehn1994applications}
VanLehn, K., Ohlsson, S., Nason, R.: Applications of simulated students: An
  exploration. Journal of artificial intelligence in education  \textbf{5},
  135--135 (1994)

\bibitem{wilson2016back}
Wilson, K.H., Karklin, Y., Han, B., Ekanadham, C.: Back to the basics:
  {Bayesian} extensions of {IRT} outperform neural networks for proficiency
  estimation. In: International Educational Data Mining Society. ERIC (2016)

\bibitem{zhang2017privbayes}
Zhang, J., Cormode, G., Procopiuc, C.M., Srivastava, D., Xiao, X.: {PrivBayes}:
  Private data release via {Bayesian} networks. ACM Transactions on Database
  Systems (TODS)  \textbf{42}(4),  1--41 (2017)

\end{thebibliography}

\end{document}